\documentclass[doublecol]{epl2}

\usepackage{graphicx}
\usepackage{dcolumn}
\usepackage{bm}
\usepackage{ulem}
\usepackage{color}
\usepackage{float}

\begin{document}

\title{Gray molasses cooling of $^{39}$K to a high phase-space density}

\author{G. Salomon \inst{1}, L. Fouch\'e \inst{1}, P. Wang \inst{1}\thanks{Present address: Institute of Opto-Electronics, Shanxi University, Taiyuan 030006, P.R.China}, A. Aspect\inst{1}, P. Bouyer\inst{2} \and T. Bourdel\inst{1}\thanks{E-mail: thomas.bourdel@institutoptique.fr}}

\shortauthor{G. Salomon, {\it et al.}}

\institute{
  \inst{1} Laboratoire Charles Fabry, Institut d'Optique, CNRS, Univ Paris-Sud, 2, Avenue Augustin Fresnel, 91127 PALAISEAU CEDEX, France\\
  \inst{2} LP2N, Univ Bordeaux 1, IOGS, CNRS, 351 cours de la Lib\'eration, 33405 Talence, France}

\date{\today}
\abstract{
We present new techniques in cooling $^{39}$K atoms using laser light close to  the D1 transition. First, a new compressed-MOT configuration is taking advantage of gray molasses type cooling induced by blue-detuned D1 light. It yields an optimized density of atoms. Then, we use pure D1 gray molasses to further cool the atoms to an ultra-low temperature of 6\,$\mu$K. The resulting phase-space density is $2\times 10^{-4}$ and will ease future experiments with ultracold potassium. As an example, we use it to directly load up to $3\times 10^7$ atoms in a far detuned optical trap, a result that opens the way to the all-optical production of potassium degenerate gases.
}

\pacs{37.10.De}{Atom cooling methods}
\pacs{37.10.Gh}{Atom traps and guides}
\pacs{67.85.-d}{Ultracold gases, trapped gases}


\maketitle

\begin{table*}[hbt]
\centering 
 \begin{small}
\begin{tabular}{ | c | c  c | c c  | c c | c c| c | c | c | c |} 
  \hline
 
Sequences & {$\delta_\textrm{\scriptsize 2C}/\Gamma$} & $I_\textrm{\scriptsize 2C}/I_\textrm{\scriptsize S}$  &   $\delta_\textrm{\scriptsize 2R}/\Gamma$  &   $I_\textrm{\scriptsize 2R}/I_\textrm{\scriptsize S}$ & $\delta_\textrm{\scriptsize 1C}/\Gamma$ & $I_\textrm{\scriptsize 1C}/I_\textrm{\scriptsize S}$  & $\delta_\textrm{\scriptsize 1R}/\Gamma$ &  $I_\textrm{\scriptsize 1R}/I_\textrm{\scriptsize S}$ & N($10^9$) &  T($\mu$K) &  n (cm$^{-3}$) & $\phi$\\ [1ex] 
\hline 
MOT with  D1 light (7\,s) & -9 & 5 & -2.7 & 3.5 & 0 & 1.5 & 0  & 0.5 & 2 & $\sim 10^4$ & - & - \\ 
D1-D2 CMOT (7\,ms)  & - & - & -1.7 & 0.5 & 3.5 & 3.5 & - & - & 1.5 & 200 & $1.3\times 10^{11}$ & $10^{-6} $\\
D1 gray molasses (7\,ms)  & - & - & - & - & 3.5 & 3.5 to 0.2  & 3.5  & 1.2 to 0.07 & 1.5  & 6 &  $1.3\times 10^{11}$ & $2\times 10^{-4}$\\ [1ex] 
\hline 

\end{tabular}
\end{small}
\caption{Experimental parameters during the various experimental sequences.  $\Gamma /2 \pi= 6\,$MHz is the natural linewidth of the excited state, $I_\textrm{\scriptsize S}=1.75\,$mW.cm$^{-2}$ is the saturation intensity.
The intensities are given at the center of a single beam. The magnetic field gradient in the MOT with D1 light and in the hybrid D1-D2 compressed-MOT(CMOT) is 10\,G.cm$^{-1}$ in the strong direction. We report the atom number $N$, the optimum temperature $T$, the density $n$ and the phase-space density $\phi$ \cite{commentphasespace}. In our setup, the different cooling steps using D1 light permit us a fiftyfold improvement in the phase-space density as compared to methods using only D2 light.}
\label{parameters} 
\end{table*}

\section{Introduction}

\begin{figure}[ht]
\centering
\includegraphics[width=0.45\textwidth]{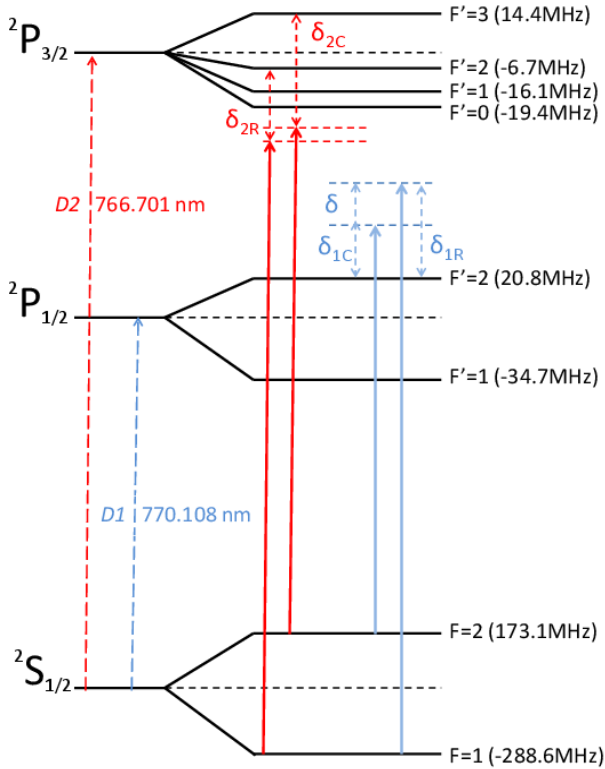}
\caption{\label{niveaux} (Color online) 
Lowest energy levels of $^{39}$K atoms. The laser detunings $\delta_{1}$ and $\delta_{2}$ relative to the D1 and D2 optical transitions are shown. The subscripts R and C refer to the repumping transitions from the $F=1$ ground state and to the cooling transitions from the $F=2$ ground state.}
\end{figure}

Ultracold atomic gases are used for the study of quantum many-body physics in strongly interacting samples \cite{Bloch08, Bloch12}. From this point of view potassium and lithium are two alkali of special interest because of the existence of both fermionic and bosonic isotopes and of broad Feshbach resonances at low magnetic fields that allow one to tune the inter-particle interactions \cite{Houbiers98, Jochim02, Khaykovich02, Strecker02, Bohn99, Loftus02, Derrico07}. These atoms however suffer from narrow hyperfine structures in their D2 excited state (see fig.\,\ref{niveaux}) preventing from efficient subdoppler cooling with light far red-detuned from the cycling transition \cite{Mewes99, Cataliotti98, Fort98}.

Several techniques have been used in order to overcome this limitation. Precise tuning of the lasers close to resonance on the D2 line has allowed one to reach subdoppler temperatures around 25\,$\mu$K in $^{39}$K \cite{Landini11, Gokhroo11}. Narrow-line laser cooling was also demonstrated for both lithium \cite{Duarte11} and potassium \cite{McKay11}. Recently, even lower temperatures were obtained in gray molasses cooling \cite{Boiron95} on the D1 transition for both lithium \cite{Grier13} and potassium \cite{Fernandes12, Nath13} (12\,$\mu$K for $^{39}$K \cite{Nath13}). In this paper, we demonstrate a yet lower temperature of 6\,$\mu$K in D1 gray molasses and  we also show that D1 light can be used to optimize previous steps of the optical cooling sequence.

Gray molasses combine two effects, $i.e.$ velocity selective coherent population trapping (VSCPT) \cite{Aspect88} and Sisyphus cooling \cite{Dalibard89}. The atoms are optically pumped in a dark state, whose departure rate varies with the square of the atom velocity, leading to less diffusion for slow atoms (VSCPT). This motional coupling to bright states happens so that the atoms repeatedly climb up the dipole potential hills in the bright states before being pumped back to the dark state (Sisyphus cooling).

The paper is organized as follows. After a description of the {\it experimental setup}, we explain how D1 light can improve the phase-space density of optically captured atoms. First, a new type of {\it hybrid D1-D2 compressed-MOT} \cite{Petrich94}  combines blue-detuned D1 light, which induces gray molasses Sisyphus cooling on the $F=2 \rightarrow F'=2$ transition, and red-detuned D2 repumping light, which gives the trapping force. It is characterized by a higher density and a much lower temperature as compared to standard compressed-MOT using only D2 light \cite{ Landini11, Gokhroo11}. We then show deep subdoppler cooling in pure {\it D1 gray optical molasses}. By ramping down the optical power, we cool all atoms to 6\,$\mu$K, far below the previously obtained temperatures in optically cooled potassium \cite{Fernandes12, Nath13}.
We finally report the direct \textit{loading of an optical dipole trap} at a phase-space density of $3\times 10^{-4}$ \cite{commentphasespace}. This result opens the route to an all optical production of a degenerate $^{39}$K quantum gas.

\section{Experimental setup}

Our experimental setup is composed of two chambers similar to those previously used for Bose-Einstein condensation of $^{87}$Rb \cite{Clement09}. A two-dimensional magneto-optical trap (2D-MOT) in the collection chamber loads a three-dimensional (3D) MOT in the science chamber. The $^{39}$K partial pressure in the collection chamber is kept at 10$^{-8}\,$mbar by heating an oven filled with metallic potassium at  100$^\circ$C and the whole chamber at 50$^\circ$C. In the science chamber the background pressure is kept at 10$^{-10}\,$mbar thanks to differential pumping. The D2 laser system is composed of a telecom fiber diode, a 2W Erbium doped fiber amplifier (Manlight), a 50/50 fiber splitter and two periodically-poled lithium nobiate waveguides (NTT Electronics) in order to produce two 100\,mW beams at 767\,nm \cite{Stern10}. Acousto-optic modulators are then used to tune the beam frequencies around the principal and repumping frequencies before recombination and injection into two 1.5$\,$W tapered amplifiers for the 2D- and 3D-MOT. 

The D1 cooling light (770$\,$nm) is produced by a semi-conductor laser diode in an interference-filter-stabilized extended cavity setup \cite{Baillard06}, which is then further amplified in a tapered amplifier. An electro-optic modulator (Qubig) is used to produce the repumping frequency. Both D1 and D2 cooling beams are then superimposed with only 10$\%$ losses using a 1$\,$nm bandwidth interference filter (Radiant Dyes Laser). The beam containing all four frequencies (needed for D1 and D2 cooling) in the same linear polarization is then sent into a 1 to 6 fiber cluster (Sh\"after-Kirchhoff). In order to produce the MOT beams, each fiber output port is then circularly polarized and collimated to a waist radius of 9$\,$mm, with a clear aperture of 24$\,$mm. This setup allows for an independent control of the D1 and D2 cooling frequencies and powers (see fig.\,\ref{niveaux}). 

\section{Hybrid D1-D2 Compressed-MOT}

Our experiment starts with loading a 3D-MOT. Surprisingly,
adding exactly resonant D1 light at this stage results in an increase of the number of trapped atoms from $1.3\times 10^9$ to $2 \times 10^9$ (see table\,\ref{parameters}). Whereas the resonant D1 light is responsible for an increase of the fluorescence rate by 25$\%$ it also causes an increase of the MOT size due to stronger photon reabsorption. As a consequence of the lower density, we observe a reduction of the rate of light assisted two body losses \cite{Prentiss88, Marcassa88} on the MOT decay. This leads to an increased number of atoms in the MOT.  

We then implement a new type of compressed-MOT. It uses both D1 light, which is blue-detuned from the $F=2 \rightarrow F'=2$ transition, and D2 repumping light, which is red-detuned from the $F=1 \rightarrow F'=2$ transition (see fig.\,\ref{niveaux} and table\,\ref{parameters}).  We observe a reduction of the rms size of the cloud from 1.4\,mm (in a D2 compressed MOT with a large detuning of the cooling beam) to 0.9\,mm, corresponding to an increase of a factor $\sim 4$ in the peak density. The temperature is then below 200\,$\mu$K as compared to 2\,mK in a D2 compressed MOT indicating a more efficient cooling process. 

Our interpretation is the following. Close to resonance from a $F=2\rightarrow F'=2$ transition and for any light polarization, there is a dark state and four bright states, which are coherent superpositions of Zeeman sublevels. Motional coupling from the dark state to bright states takes place due to polarization gradients in our 3D $\sigma^+  -\sigma^-$ configuration. As a consequence, a gray molasses type cooling occurs for D1 light blue-detuned from the $F=2 \rightarrow F'=2$ transition \cite{Fernandes12}. The magnetic field of the order of one gauss in the trapping region is not high enough to immediately destroy the coherence of the dark state and allows Sisyphus cooling cycles to take place.  This explains our observed low temperature in the compressed-MOT. Red-detuned D2 light is used to repump the atoms. In combination with the magnetic field gradient, it also leads to the only trapping force in our hybrid compressed-MOT. Furthermore, we have observed that better compression is obtained when reducing the repump intensity below the saturation intensity (see table \,\ref{parameters}). This is crucial in order to reduce photon reabsorption, a process which limits the density.
For the optimal parameters, we observe a fluorescence rate in our hybrid D1-D2 compressed MOT that is five times lower than in our MOT \cite{fraction}. 

\section{D1 gray optical molasses}
For the last cooling step, we use bichromatic beams in a $\Lambda$-configuration from the two $F=1$ and $F=2$ hyperfine ground states to the blue side of the D1 line, a setup which permits deep subdoppler cooling \cite{Grier13, Fernandes12, Nath13}. This technique is first taking advantage of the gray molasses cooling mechanism on the D1 $F=2 \rightarrow F'=2$ transition, which is also at play in our compressed-MOT. Moreover, the two light fields at the Raman condition $\delta_{1 \textrm{\scriptsize C}}=\delta_{1 \textrm{\scriptsize R}}=\Delta$ introduce new dark states which are coherent superpositions of the $F=1$ and $F=2$ sublevels as in electromagnetically induced transparency \cite{Arimondo96}. There are thus additional possible gray molasses cooling cycles, which lead to even lower temperatures. Experimentally, we have found that a repumping to cooling intensity ratio of one third is optimal \cite{Nath13}. 

\begin{figure}[ht]
\centering
\includegraphics[width=0.49\textwidth]{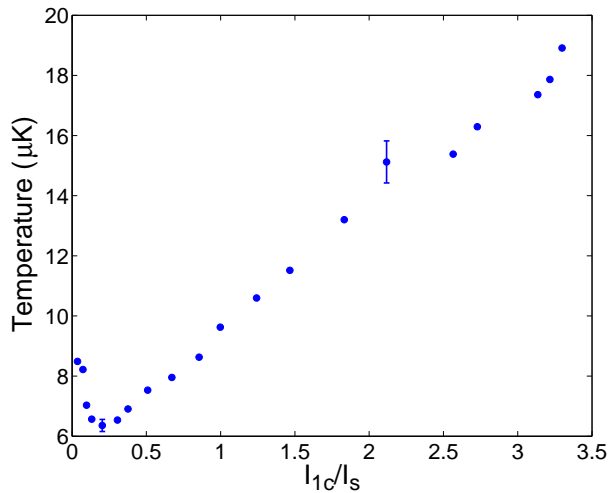}
\caption{\label{intensity} (Color online) 
Temperature of the D1 gray molasses as a function of the D1 peak intensity per beam. The D1 repumping intensity corresponds to one third of the cooling intensity.}
\end{figure}

We first set $\delta_{1 \textrm{\scriptsize C}}=\delta_{1 \textrm{\scriptsize R}}=\Delta=2\pi \times 21\,$MHz$=3.5\,\Gamma$ and study the influence of the D1 light intensity. With a D1 peak intensity per beams of 3.5\,$I_\textrm{\scriptsize S}$, all atoms from the compressed MOT are captured by the optical molasses and cooled below 25\,$\mu$K in less than 2 ms. This shows the great cooling efficiency and the large capture range of the gray molasses technique. We thus do not need a precooling stage using D2 molasses. In order to reach lower temperatures, we then ramp down the  D1 power linearly in 7\,ms. During this time, the atom expansion is negligible compared to the initial size of the compressed-MOT cloud. In fig.\,\ref{intensity}, we present the final temperature as a function of the final D1 molasses power. We find that the temperature decreases linearly with the final optical power for sufficiently high intensity. This may be understood, in analogy with the behavior of bright molasses, from the reduction of the induced light shifts in the dressed state picture \cite{Grier13}. For a cooling laser intensity of 0.2\,$I_\textrm{\scriptsize S}$, we find a minimal temperature of $\sim 6\,\mu$K which is a factor two below previously achieved temperatures \cite{Nath13}. If we start directly with this low intensity, the capture velocity of the gray molasses is reduced and only 10$\%$ of the atoms from the hybrid D1-D2 compressed-MOT are efficiently cooled (only $\sim 1.5 \%$ from a D2 compressed-MOT at 2\,mK). These are cooled to the same temperature of 6\,$\mu$K showing the absence of density effect. The others are lost, revealing the importance of the power ramp to dynamically tune the velocity capture range. At lower intensity, the cooling mechanism is not efficient enough and the observed temperature increases.

\begin{figure}[ht]
\centering
\includegraphics[width=0.49\textwidth]{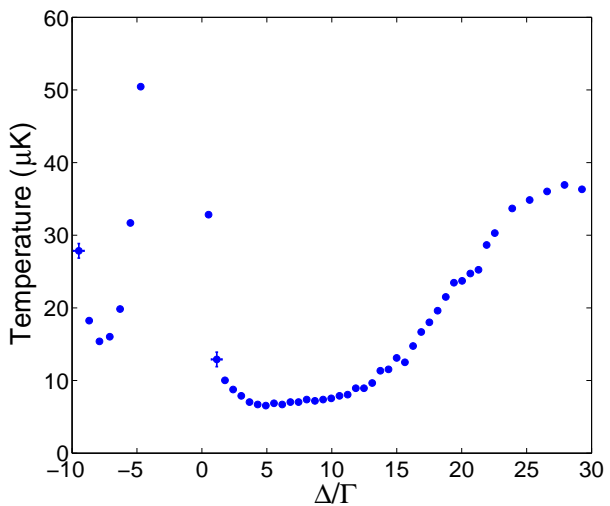}
\caption{\label{detuning} (Color online) 
Temperature of the D1 gray molasses as a function of the global detuning $\Delta$ of the D1 cooling beams. $\Delta=0$ corresponds to being on resonance with the D1 $F'=2$ state.  $\Delta=-9.3\,\Gamma$ corresponds to being on resonance with the D1 $F'=1$ state.}
\end{figure}

In figure \ref{detuning}, we then study the temperature as a function of the global detuning $\Delta$ of the D1 cooling beams. For positive detunings, we find that cooling works over a broad range. The cooling is found to work best when the detuning $\Delta=\delta_\textrm{\scriptsize 1C}=\delta_\textrm{\scriptsize 1R}$ is between 3.5\,$\Gamma$ and 7$\,\Gamma$. For negative detunings, cooling also occurs  around $\Delta=-7.5\,\Gamma$. We interpret this as gray molasses working with the $F'=1$ state. However the minimum temperature obtained at this detuning while optimizing the molasses parameters was about 10\,$\mu$K. Being too close and red-detuned from the $F'=2$ state leads to heating \cite{Grier13}. This is confirmed by the fact that red-detuned from the $F'=1$ state (and thus also from the $F'=2$ state), no cooling is observed. 

\begin{figure}[ht]
\centering
\includegraphics[width=0.49\textwidth]{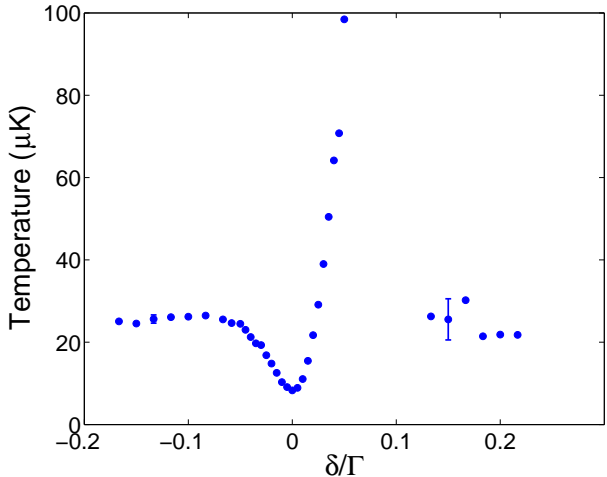}
\caption{\label{raman} (Color online) 
Temperature of the D1 gray molasses as a function of Raman detuning $\delta$. Deep cooling is found at the Raman resonance. The cooling at 25 $\mu$K observed on both sides of the resonance is attributed to the gray molasses mechanism working on the $F=2\rightarrow F'=2$ transition. }
\end{figure}

Finally, we have also studied the dependance of the final temperature as a function of the Raman detuning $\delta=\delta_\textrm{\scriptsize 1R}-\delta_\textrm{\scriptsize 1C}$ (see fig.\,\ref{raman}). We observe a narrow feature going from deep cooling at $\delta=0$ to relative heating at $\delta\approx 2\pi\times 0.4\,$MHz, which is characteristic of the gray molasses Sisyphus mechanism involving the dark states from the $\Lambda$ configuration \cite{Grier13, Nath13}. Maximum heating is observed when the typical light shift of the bright states corresponds to the Raman detuning such that the atoms are maximally coupled to bright states at the top of the potential hills. Actually, our very narrow feature in $\delta$, as compared to other experiments \cite{Grier13, Nath13}, is correlated to a strongly reduced power at the end of the D1 gray molasses. 

In addition to the sharp resonant feature close to $\delta=0$, we observe a background cooling at 25\,$\mu$K, which we interpret as gray molasses cooling in the $F=2\rightarrow F'=2$ manifold without hyperfine coherence. Whereas all the atoms are cooled for $\delta<0$, only 25$\%$ of the cloud is captured by this mechanism for $\delta>0$ due to the competition between heating and cooling processes.  We observe that for $\delta>0$, groups of atoms tend to accumulate at additional non-zero velocities proportional to $\delta$. For each of these velocities, the Doppler effect shifts the Raman transition into resonance for a pair of perpendicular beams. This observation directly shows the importance of the Raman dark state cooling in our system. Experimentally, another way of reducing the coherence of the dark states is to add a magnetic field. We find a very high sensitivity as a function of magnetic field with the temperature increasing as $\Delta T \sim 300(100)\,\mu$K/G$^2$. Even with such a high sensitivity, it is unlikely that uncontrolled stray magnetic fields are playing a role in our observed minimal temperature. 

\section{Loading of an optical trap}

\begin{figure}[ht]
\centering
\includegraphics[width=0.4\textwidth]{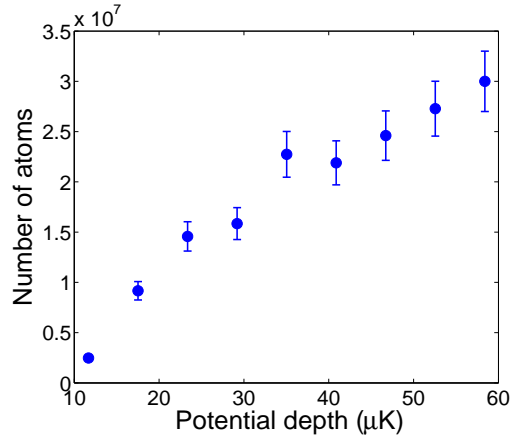}
\caption{\label{loading} (Color online) 
Number of atoms trapped in the optical dipole trap as a function of the optical trap depth.}
\end{figure}

When loading an optical dipole trap, D1 gray molasses technique has several advantages. As compared to subdoppler cooling using the D2 transition \cite{Landini11}, it permits us to reach a lower temperature (6\,$\mu$K  compared to 25\,$\mu$K) and a higher phase-space density ($2\times 10^{-4}$ \cite{commentphasespace} compared to $1.5\times 10^{-5}$ in \cite{Landini11}). These two ingredients are crucial for efficient loading of an optical trap. In addition,  we expect the trapping laser not to disturb the gray molasses because its induced light shift affects the rather insensitive global detuning $\Delta$ of the D1 molasses and not the critical Raman detuning $\delta$. Experimentally, we focus a single trapping beam at 1550\,nm down to 140\,$\mu$m on the molasses. We then switch off the molasses beams and keep the optical trap on during 80 ms to let the radially untrapped atoms fall under gravity (see fig.\,\ref{loading}). The observed number of trapped atoms decreases with the laser intensity or equivalently with the trap depth. This is expected from a larger region with a light-shift potential larger than the molasses temperature. At low potential depth (below 20\,$\mu$K), gravity is not any more negligible and the actual trap depth in the vertical direction is further reduced (eventually to $\sim$4\,$\mu$K, below the molasses temperature, for a potential depth of 12\,$\mu$K) explaining a lower trapped atom number. For the maximum power of 22\,W in the trapping beam, $3\times 10^7$ atoms are captured at a temperature of 18\,$\mu$K. The phase-space density at this stage is $3\times 10^{-4}$, which corresponds to a fifteen-fold gain as compared to the value we had in a similar configuration in a rubidium experiment producing Bose-Einstein condensates with $1.5\times10^5$ atoms \cite{Clement09}.

\section{Conclusion}

In conclusion, we show that using D1 cooling beams in addition to D2 cooling beams in experiments with potassium permits us to increase the number of atoms in the MOT, to achieve a higher density in a new type of hybrid D1-D2 compressed-MOT and to greatly reduce the temperature to 6\,$\mu$K in $\Lambda$-enhanced D1 optical molasses. Although some questions remain for example about the origin of the minimal temperature in D1 gray molasses, the obtained phase-space density up to $2\times 10^{-4}$ in an optically cooled sample will certainly ease future ultracold potassium experiments. Moreover, our methods are quite general and are likely to be efficient for a variety of atomic species including those with a narrow D2 hyperfine structure. Direct transfer from the optically cooled sample to an optical trap can then be made quite efficient and all optical cooling of potassium atoms to quantum degeneracy should be possible, simplifying existing techniques using a mixture \cite{Modugno01, Roati07} or an intermediate magnetic trap \cite{Landini12}. D1 gray molasses could also be useful to extend single atom fluorescence imaging \cite{Bakr09, Sherson10} to lithium or potassium. 

\section{Acknowledgements}

We acknowledge F. Moron, A. Villing, and O. Lesage for technical assistance, F. Chevy, C.\,S. Unnikrishnan for discussions. This research was supported by CNRS, Minist\`ere de l'Enseignement Sup\'erieur et de la Recherche, Direction G\'en\'erale de l'Armement, ANR-12-BS04-0022-01, RTRA: Triangle de la physique, iSense, ERC senior grant Quantatop. LCFIO is member of IFRAF.


\begin{thebibliography}{00}


\bibliographystyle{apsrev4-1}



\bibitem{Bloch08}
Bloch I., Dalibard J., Zwerger W., Rev. Mod. Phys. {\bf 80} (2008) 885.

\bibitem{Bloch12}
Bloch I., Dalibard J., Nacimb\`ene S., Nature Physics {\bf 8} (2012) 267 .

\bibitem{Houbiers98}
Houbiers M., Stoof H.T.C., McAlexander W. I., and Hulet R. G.  Phys. Rev. A {\bf 57} (1988) R1497 .

\bibitem{Jochim02}
Jochim S., Bartenstein M., Hendl G., Hecker Denschlag J., Grimm R., Mosk A., and Weidem\"uller W., Phys. Rev. Lett. {\bf 89}  (2002) 273202.

\bibitem{Khaykovich02}
Khaykovich L., Schreck F., Ferrari G., Bourdel T., Cubizolles J., Carr L. D., Castin Y., and Salomon C., Science {\bf 296} (2002) 1290.

\bibitem{Strecker02}
Strecker K.\,E., Partridge G. B., Truscott A. G., and Hulet R. G., Nature {\bf 417}  (2002) 150.

\bibitem{Bohn99}
Bohn J.\,L., Burke J. P., Greene C. H., Wang H., Gould P. L., and Stwalley W. C., Phys. Rev. A {\bf 59} (1999)  3660.

\bibitem{Loftus02}
Loftus T., Regal C. A., Ticknor C., Bohn J. L., and Jin D. S., Phys. Rev. Lett. {\bf 88} (2002) 173201.

\bibitem{Derrico07}
D'Errico C., Zaccanti M., Fattori M., Roati G., Inguscio M., Modugno G., and Simoni A.,  New J. Phys. {\bf 9} (2007) 223.

\bibitem{Mewes99}
Mewes M.-O., Ferrari G., Schreck F., Sinatra A., and Salomon C., Phys. Rev. A {\bf 61}  (1999) 011403(R).

\bibitem{Cataliotti98}
Modugno G., Benk C., Hannaford P., Roati G., and Inguscio M., Phys. Rev. A {\bf 60}  (1999) R3373; Cataliotti F. S., Cornell E. A., Fort C., Inguscio M., Marin F., Prevedelli M., Ricci L., and Tino G. M., ibid. {\bf 57}  (1998) 1136.

\bibitem{Fort98}
FortC. , Bambini A., Cacciapuoti L., Cataliotti F. S., Prevedelli M., Tino G. M. and Inguscio M., Eur. Phys. J. D {\bf 3}  (1998) 113.

\bibitem{Landini11}
Landini M., Roy S., Carcagní L., Trypogeorgos D., Fattori M., Inguscio M., and Modugno G., Phys. Rev. A {\bf 84}  (2011) 043432.

\bibitem{Gokhroo11}
Gokhroo V., Rajalakshmi G., Easwaran R. K., and Unnikrishnan C. S., J. Phys. B {\bf 44} (2011) 115307.

\bibitem{Duarte11}
Duarte P. M., Hart R. A., Hitchcock J. M., Corcovilos T. A., Yang T.-L., Reed A., and Hulet R. G., Phys. Rev. A {\bf 84} (2011) 061406(R).

\bibitem{McKay11}
McKay D. C., Jervis D., Fine D. J., Simpson-Porco J. W., Edge G. J. A., and Thywissen J. H., Phys. Rev. A {\bf 84} (2011)  063420.

\bibitem{Boiron95}
Boiron D., Triche C., Meacher D. R., Verkerk P., and Grynberg G., Phys. Rev. A {\bf 52} (1995) R3425.

\bibitem{Grier13}
Grier A.T., Ferrier-Barbut I., Rem B.S., Delehaye M., Khaykovich L., Chevy F., and Salomon C., Phys. Rev. A {\bf 87} (2013) 063411.

\bibitem{Fernandes12}
Rio Fernandes D., Sievers F., Kretzschmar N., Wu S., Salomon C. and Chevy F., Euro. Phys. Lett.  {\bf 100} (2012)  63001.

\bibitem{Nath13}
Nath D., Easwaran R.\,K., Rajalakshmi G., Unnikrishnan C.\,S., arXiv:1305.5480 preprint, 2013.

\bibitem{Aspect88}
Aspect A., Arimondo E., Kaiser R., Vansteenkistet N., and Cohen-Tannoudji C., Phys. Rev. Lett. {\bf 61} (1988) 826.

\bibitem{Dalibard89}
Dalibard J. and Cohen-Tannoudji C., J. Opt. Soc. Am. B {\bf 6} (1989) 2023.

\bibitem{Petrich94}
Petrich W., Anderson M.\,H., Ensher J.\,R., Cornell E.\,A.,
J. Opt. Soc. Am. B {\bf 11} (1994) 1332.

\bibitem{commentphasespace}
The given phase-space density does not take into account the different internal states.

\bibitem{Clement09}
Cl\'ement J.-F., Brantut J.-P., Robert-de-Saint-Vincent M., Nyman R. A., Aspect A., Bourdel T., and Bouyer P., Phys. Rev. A {\bf 79}  (2009) 061406(R).

\bibitem{Stern10}
Stern G., Allard B., Robert-de-Saint-Vincent M., Brantut J.-P., Battelier B., Bourdel T., Bouyer P., applied optics {\bf 49} (2010) 1.

\bibitem{Baillard06}
Baillard X., Gauguet A., Bize S., Lemonde P., Laurent P., Clairon A., and Rosenbusch P., Opt. Comm. {\bf 266} (2006)  609.


\bibitem{Prentiss88}
Prentiss M., Cable A., Bjorkholm J. E., Chu S., Raab E. L., and Pritchard D. E., Opt. Lett. {\bf 13} (1988) 452.

\bibitem{Marcassa88}
Marcassa L., Bagnato V., Wang Y., Tsao C., Weiner J., Dulieu O., Band Y. B., and Julienne P. S., Phys. Rev. A {\bf 47} (1993) R4563.

\bibitem{fraction}
 40\,\% (60\,\%) of atoms are in the F=2 (F=1) manifold at the optimal parameters. 
 
 \bibitem{Arimondo96}
Arimondo E., in {\it Progress in Optics}, edited by E. Wolf {\bf 35}  (Elsevier, Amsterdam) (1996) 257 .

\bibitem{Modugno01}
Modugno G., Ferrari G., Roati G., Brecha R. J., Simoni A., and Inguscio M., Science {\bf 294} (2001) 1320.

\bibitem{Roati07}
Roati G., Zaccanti M., D'Errico C., Catani J., Modugno M., Simoni A., Inguscio M., and Modugno G., Phys. Rev. Lett. {\bf 99} (2007) 010403.

\bibitem{Landini12}
Landini M., Roy S., Roati G., Simoni A., Inguscio M., Modugno G., and Fattori M., Phys. Rev. A {\bf 86} (2012) 033421.

\bibitem{Bakr09}
Bakr W. S., Gillen J. I., Peng A., Tai M. E., F\"olling S., Greiner M., Nature {\bf 462} (2009) 74 .

\bibitem{Sherson10}
Sherson J. F., Weitenberg C., Endres M., Cheneau M., Bloch I., Kuhr S.,
Nature {\bf 467} (2010) 68.

\end{thebibliography}
\end{document}